\documentclass[pra,twocolumn,showpacs,preprintnumbers,amsmath,amssymb,superscriptaddress,longbibliography]{revtex4-2}

\usepackage{graphicx}
\usepackage{dcolumn}
\usepackage{bm}
\usepackage{color}
\usepackage{bbm}
\usepackage[colorlinks=true,citecolor=blue,linkcolor=blue,urlcolor=blue]{hyperref}
\usepackage{appendix}

\begin{document}
\title{Tunable quantum router with giant atoms, implementing quantum gates, teleportation, non-reciprocity,
 and circulators}
\author{Rui-Yang Gong}
\thanks{These two authors contributed equally.}
\affiliation{School of Physics, Sun Yat-sen University, Guangzhou 510275, China}
\author{Zi-Yu He}
\thanks{These two authors contributed equally.}
\affiliation{School of Physics, Sun Yat-sen University, Guangzhou 510275, China}
\author{Cheng-He Yu}
\affiliation{State Key Laboratory for Mesoscopic Physics, School of Physics, Frontiers Science Center for Nano-optoelectronics, $\&$ Collaborative Innovation Center of Quantum Matter, Peking University, Beijing 100871, China}
\affiliation{Hefei National Laboratory, Hefei 230088, China}
\author{Ge-Fei Zhang}
\affiliation{School of Physics, Sun Yat-sen University, Guangzhou 510275, China}
\author{Franco Nori}
\affiliation{Theoretical Quantum Physics Laboratory, Cluster for Pioneering Research, RIKEN, Wakoshi, Saitama 351-0198, Japan}
\affiliation{Center for Quantum Computing, RIKEN, Wakoshi, Saitama 351-0198, Japan}
\affiliation{Department of Physics, University of Michigan, Ann Arbor, Michigan 48109-1040, USA}
\author{Ze-Liang Xiang}
\email{xiangzliang@mail.sysu.edu.cn}
\affiliation{School of Physics, Sun Yat-sen University, Guangzhou 510275, China}
\date{\today}

\begin{abstract}
The unique photon-scattering phenomena of giant-atom systems offer a novel paradigm for exploring innovative quantum optics phenomena and applications. Here, we investigate a giant-atom configuration embedded in a dual-rail waveguide, whose scattering behavior is analytically derived based on a four-port model and affected by both waveguide-induced and interatomic interaction phases. One can modulate these phases to achieve targeted routing and non-reciprocal scattering of photons. Furthermore, using such a configuration, we propose quantum applications such as quantum storage, path-encoded quantum gates (e.g., CNOT gate), quantum teleportation, and quantum circulators. This configuration can be implemented with state-of-the-art solid-state quantum systems, enabling a wide range of quantum applications and facilitating the development of quantum networks.
\end{abstract}
\maketitle

\section{Introduction}
Quantum networks are considered the most promising framework for scalable quantum information processing and communication~\cite{kimble_quantum_2008,doi:10.1126/science.aam9288}, where quantum state transfer as a key ingredient has attracted increasing attention in recent years~\cite{PhysRevX.7.011035,xiangPhysRevLettUniversalTime-DependentControlScheme2023,PhysRevLett.131.250801}.
With the growth of qubit numbers and quantum network size, targeted routing of quantum states between different nodes has become necessary in realizing large-scale quantum information processing~\cite{GU2017Microwavephotonicswithsuperconductingquantumcircuits,2023reviewApplicationsofsinglephotons}.
In most works, targeted routing is achieved using external drivers to counteract Lorentz reciprocity via the Faraday effect~\cite{2018PhysRevAppliedElectromagneticNonreciprocity,1953RevModPhysTheFerromagneticFaradayEffect}. 
However, these devices are bulky, lossy, and require extra pumps, making them unsuitable for on-chip integration~\cite{PhysRevA.94.063817Nonreciprocalfew-photonrouting,PhysRevA.98.063809Phase-modulatedsingle-photonrouter}. 
To address these challenges, various approaches have been proposed in recent years~\cite{PhysRevA.80.062109,PhysRevA.98.022310,science.1257671routingoptic,PhysRevX.3.031013All-OpticalRouter,PhysRevA.102.011502,Cai_2024}. Among them, waveguide-QED systems are widely employed, including photonic crystal waveguides~\cite{RevModPhys.90.031002,Gonzalez-Tudela2024,yanTargetedPhotonicRouters2018}, coupled resonators~\cite{PhysRevA.78.063827,PhysRevLett.101.100501,PhysRevLett.111.103604Single-PhotonRouter2,PhysRevA.89.013805}, optical cavities~\cite{science.1254699All-opticalroutingofsinglphotons1,PhysRevA.99.033827}, optomechanical waveguides~\cite{PhysRevA.85.021801Optomechanical}, or superconducting circuit waveguides~\cite{PhysRevLett.107.Single-PhotonRouter1,PhysRevLett.111.063601CQEDrouter,2021PhysRevAppliedQuantumRouterSCQubit}.

Generally, atoms can be regarded as points because of their negligible sizes compared to the wavelength of the field~\cite{royColloquiumStronglyInteracting2017}. This has also been widely adopted in atom-embedded waveguide-QED systems\cite{PhysRevLett.106.020501,Chang_2012,PhysRevA.88.043806,PhysRevLett.115.163603,doi:10.1126/science.1244324,PhysRevA.97.063809,Mirhosseini2019}. Recently, superconducting qubits have been coupled to both surface acoustic waves and microwave waveguides at multiple spatially separated points across considerable distances~\cite{gustafsson2014propagating,kannanWaveguideQuantumElectrodynamics2020,2023PhysRevXResonanceFluorescenceofaChiralArtificialAtom,PhysRevA.103.023710}. These facts challenge the widely acknowledged dipole approximation in quantum optics~\cite{scully1999quantumoptics,Agarwal_2012}, representing a new paradigm in quantum optics called giant atoms~\cite{10.1007/978-981-15-5191-8_12}. In such systems, novel phenomena can be observed by engineering the geometric structure of coupling points, such as decoherence-free interatomic interaction~\cite{kockumDecoherenceFreeInteractionGiant2018,carolloMechanismDecoherencefreeCoupling2020,soroChiralQuantumOptics2022,duComplexDecoherencefreeInteractions2023,PhysRevA.110.052612}, non-exponential atomic decay~\cite{anderssonNonexponentialDecayGiant2019}, bound states in the continuum~\cite{guoGiantAcousticAtom2017,GUOPhysRevResearchOscillatingboundstates,wangTunableChiralBound2021,xiaoBoundStateGiant2022}, modiﬁed topological effects~\cite{PhysRevA.104.053522topoga1,PhysRevA.106.033522topoga2,zhu2024single,PhysRevResearch.5.023031,PhysRevX.11.011015}, sudden birth of entanglement~\cite{PhysRevA.108.023728,PhysRevA.106.063703,GenerationofMaximallyEntangled2023}, frequency-dependent relaxation rates~\cite{friskkockumDesigningFrequencydependentRelaxation2014}, and Lamb shifts~\cite{PhysRevA.103.023710}. 
These phenomena are caused by interference effects among multiple paths introduced by multi-coupling points.
In solid-state systems, superconducting qubits have been fabricated with multi-coupling points as a giant artificial molecule to realize the on-demand control of flying microwave photons~\cite{almanakly2024deterministic,kannan2023demand}, heralding the potential for engineering all-to-all quantum communication and large-scale quantum networks~\cite{guimondUnidirectionalOnchipPhotonic2020a,wangChiralQuantumNetwork2022a,doi:10.1126/sciadv.aba4935}.
However, as future quantum networks become more complex and functional, their components, such as on-chip routers, will require more efficient and simplified designs. 
\begin{figure*}[htpb]
    \centering
    \includegraphics[width=0.85\linewidth]{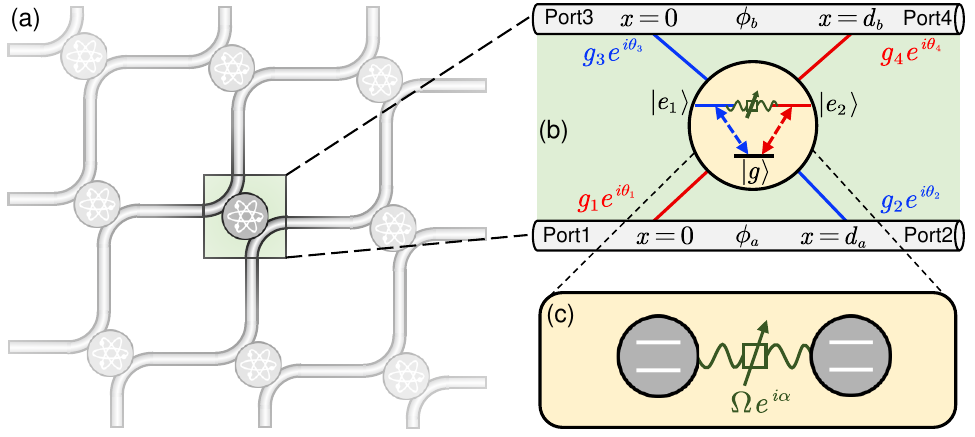}
    \caption{Schematic of a giant-atom node model in a dual-rail waveguide network. (a) Overview of the network with giant-atom nodes at waveguide intersections. (b) The effective giant atom coupled to the waveguide, with different coupling strengths and phases. (c) Internal structure of the giant atom, consisting of two interacting qubits.}
    \label{fig:1}
\end{figure*}

In this work, we propose a tunable four-port quantum router by embedding a giant-atom node in a dual-rail waveguide with multiple coupling points. We investigate the effects of interatomic interactions and the spatial separation between coupling points on single-photon scattering of the giant atom. Surprisingly, the model can exhibit chirality of transport modulated by the coupling phase. With this property, a tunable targeted router is constructed based on the multi-path interference. Here, we discuss targeted routing in two different regimes, depending on their functions. One is the trans-waveguide regime, where photons can be completely transferred from the lower waveguide to the upper waveguide. The other is the unidirectional regime, which enables the unidirectional transmission of flying photons and shows anti-reflection scattering characteristics. Ideally, these routing approaches can achieve up to 100\% efficiency, thus allowing precise on-demand control of flying photons. Then, we use the unidirectional transmission of the router to realize \textit{path-encoded quantum gates} and \textit{quantum teleportation}. In addition, a \textit{tunable circulator} is studied based on the property of \textit{non-reciprocal} scattering under photon detuning. Finally, we demonstrate the \textit{robustness} of the system against atomic decay and parameter mismatches under feasible experimental conditions. These results highlight the significant potential of our system for practical application in quantum networks.


\section{model}
\label{model and methods}

We consider a giant-atom waveguide-QED system, where the giant atom couples to a dual-rail waveguide in separate points, as depicted in Figs.~\ref{fig:1}(a) and \ref{fig:1}(b). This giant atom, with a $\nabla$-type structure, can be regarded as two coupled qubits with coupling strength $\Omega$ and phase $\alpha$, as shown in Fig.~\ref{fig:1}(c). 
The transition $|g\rangle \leftrightarrow |e_1\rangle $ of the $\nabla$-type atom is coupled to the lower waveguide $W_a$ at the point $x=d_a$ and the upper waveguide $W_b$ at the point $x=0$, respectively. The other transition $|g\rangle \leftrightarrow |e_2\rangle $ of the $\nabla$-type atoms is coupled to the lower waveguide $W_a$ at the point $x=0$ and the upper waveguide $W_b$ at the point $x=d_b$, respectively, where $|e_1\rangle, |e_2\rangle $ and $|g\rangle $ are two different excited states $|10\rangle,|01\rangle$ and ground state $|00\rangle$ of two coupled qubits~\cite{roushan2017chiral}.

The total Hamiltonian of the system under the linearized waveguide dispersion approximation \cite{shenTheorySinglephotonTransport2009b,shenTheorySinglephotonTransport2009a} can be expressed as ($\hbar=1$)
\begin{equation}
    	H=H_{w}+H_a+H_I,
\end{equation}
where
\begin{align}
	H_{w}=&\,\sum_{l=L,R}{\int_{-\infty}^{+\infty}{d}x\left[ a_{l}^{\dagger}(x)\left( \omega _0+if_lv_g\frac{\partial}{\partial x} \right) a_l(x) \right.}\nonumber\\
	&\,+\left. b_{l}^{\dagger}(x)\left( \omega _0+if_lv_g\frac{\partial}{\partial x} \right) b_l(x) \right],\\
	H_a=&\,\sum_{i=1,2}{\left( \omega _{e_i}-i\frac{\gamma _{e_i}}{2} \right) \vert e_i\rangle \langle e_i \vert} \nonumber\\
        &\, +\left( \Omega e^{i\alpha}\vert e_1 \rangle \langle e_2 \vert+\mathrm{H.c.} \right),\\
    H_I=&\,\int_{-\infty}^{+\infty}{d}x\left\{ \delta (x)g_1e^{i\theta _1}\left[ a_{R}^{\dagger}(x)+a_{L}^{\dagger}(x) \right] |g\rangle \left\langle e_2 \right| \right.\nonumber\\
	&\,+\delta (x-d_a)g_2e^{i\theta _2}\left[ a_{R}^{\dagger}(x)+a_{L}^{\dagger}(x) \right] |g\rangle \langle e_1 \vert\nonumber\\
	&\,+\delta (x-d_b)g_4e^{i\theta _4}\left[ b_{R}^{\dagger}(x)+b_{L}^{\dagger}(x) \right] |g\rangle \left. \langle e_2 \right|\nonumber\\
	&\,+\delta (x)g_3e^{i\theta _3}\left[ b_{R}^{\dagger}(x)+b_{L}^{\dagger}(x) \right] |g\rangle \left. \langle e_1 \right|+\left. \mathrm{H}.\mathrm{c}. \right\}.
\end{align}
$H_w$ is the Hamiltonian of the dual-waveguide modes with group velocity $v_g$ and central frequency of the linearized dispersion $\omega_0$, where $f_L=1$ and $f_R=-1$, $a_{R,L}^{\dagger}/b_{R,L}^{\dagger}(a_{R,L}/b_{R,L})$ are the creation (annihilation) operators of the rightward-propagating and leftward-propagating photons in the lower waveguide $W_a$ and upper waveguide $W_b$, respectively. $H_a$ is the effective Hamiltonian of the $\nabla$-type giant atom, where $\omega_{e_i}$ and $\gamma_{e_i}$ describe the transition frequency and the dissipation rate between the excited state $\vert e_i \rangle $ and the ground state $\vert g \rangle $, respectively. 
$H_I$ describes the interaction between the giant atom and a dual-rail waveguide at the multi-coupling points, where $\delta(x)$, $\delta(x-d_a)$, and $\delta(x-d_b)$ indicate that the coupling of the giant atoms to the upper and lower waveguide occurs at $x=0$, $x=d_a$, and $x=d_b$ of the spatial coordinates.

Two different transitions $|g\rangle \leftrightarrow |e_{1,2}\rangle$ between the giant atom and waveguides in $H_{I}$ are illustrated by the blue and red lines in Fig.~\ref{fig:1}(b). The transition $|g\rangle \leftrightarrow |e_{1}\rangle $ of the frequency $\omega_{e_1}$ is coupled to the lower waveguide $W_a$ with the coupling strength $g_2e^{i\theta _2}$ and the upper waveguide $W_b$ with the coupling strength $g_3e^{i\theta _3}$ at the coupling points $x=d_a$ and $x=0$, respectively. Similarly, the transition $|g\rangle \leftrightarrow |e_{2}\rangle $ of the frequency $\omega_{e_2}$ is coupled to the upper waveguide $W_b$ with the coupling strength $g_4e^{i\theta _4}$ and the lower waveguide $W_a$ with the coupling strength $g_1e^{i\theta _1}$ at the coupling points $x=d_b$ and $x=0$, respectively. The state of the $\nabla$-type atom and dual-rail waveguide in the single-excitation subspace can be written as
\begin{equation}
\label{wavefuncgeneral}
  \begin{aligned}
     |\Psi \rangle =&\sum_{l=L,R}{\int_{-\infty}^{+\infty}{d}x\left[ \Phi _{al}(x)a_{l}^{\dagger}(x) \right.}\left. +\Phi _{bl}(x)b_{l}^{\dagger}(x) \right] |0,g\rangle 
    \\ &+\sum_{i=1,2}{u_{e_i}|0,e_i\rangle},
  \end{aligned}
\end{equation}
where $|0,g\rangle$ represents atoms in the ground state with zero photons in the waveguides. $\Phi_{bl}(x)$[$\Phi _{al}(x)$] is the wavefunction of the leftward and rightward propagating photons in the upper (lower) waveguide. In addition, $u_{e_n}$ is the amplitude of the $\nabla$-type giant atom in different excited states when there are no photons in the waveguide due to the confinement of the single-exciton subspace. Without losing generality, we assume that the photon is injected in port 1. According to the Bethe ansatz~\cite{shenTheorySinglephotonTransport2009a,shenTheorySinglephotonTransport2009b}, the series of the wave function of the propagating photons is
\begin{equation}
\label{S14 wavefunction}
    \begin{aligned}
	\Phi _{aR}(x)=&e^{ik_ax}\left\{ \Theta (-x)+W_{a}^{R}\left[ \Theta (x)-\Theta \left( x-d_a \right) \right] \right.  
	\\
	&\left.+s_{1\rightarrow 2}\Theta \left( x-d_a \right) \right\},\\
	\Phi _{aL}(x)=&e^{-ik_ax}\left\{ s_{1\rightarrow 1}\Theta (-x)+W_{a}^{L}\left[ \Theta (x)-\Theta \left( x-d_a \right) \right] \right\},
 \\
	\Phi _{bR}(x)=&e^{ik_bx}\left\{ W_{b}^{R}\left[ \Theta (x)-\Theta \left( x-d_b \right) \right] +s_{1\rightarrow 4}\Theta \left( x-d_b \right) \right\},\\
	\Phi _{bL}(x)=&e^{-ik_bx}\left\{ s_{1\rightarrow 3}\Theta (-x)+W_{b}^{L}\left[ \Theta (x)-\Theta \left( x-d_b \right) \right] \right\},\\
\end{aligned}
\end{equation}
where $\Theta (x)$ is the Heaviside step function, $s_{n\rightarrow m},n,m\in \left\{ 1,2,3,4 \right\}$ are the scattering coefficients of the four-port model, representing the probability amplitude of a free propagating photon incident at port $n$ and scattered to port $m$ after interacting with the giant atom at four coupling points, respectively. The scattering coefficients $W_{a}^{R}$ and $W_{a}^{L}$ $(W_{b}^{R}$ and $W_{b}^{L})$ describe the rightward and leftward photon transmission, respectively, in the region of $0<x<d_a$ ($0<x<d_b$) between the coupling points in the lower (upper) waveguide. The photon scattering amplitudes $s_{1 \rightarrow n}$ can be solved by inserting Eqs.~(\ref{wavefuncgeneral}) and (\ref{S14 wavefunction}) into the Schr\"{o}inger equation $H|\Psi\rangle=E|\Psi\rangle$, with details shown in Appendix~\ref{appendix:Solution of the scattering matrix}. Different from previous works~ \cite{chenNonreciprocalChiralSinglephoton2022b,zhuSpatialnonlocalityinducedNonMarkovianElectromagnetically2022a,zhouChiralNonreciprocalSinglephoton2023}, we assume that the two waveguides are identical, where the propagating photon wavevectors satisfy the relation $k_{a,b}=(E-\omega_0)/v_g$ in waveguide $W_{a,b}$.

\begin{figure}[htpb]
    \centering
    \includegraphics[width=1\linewidth]{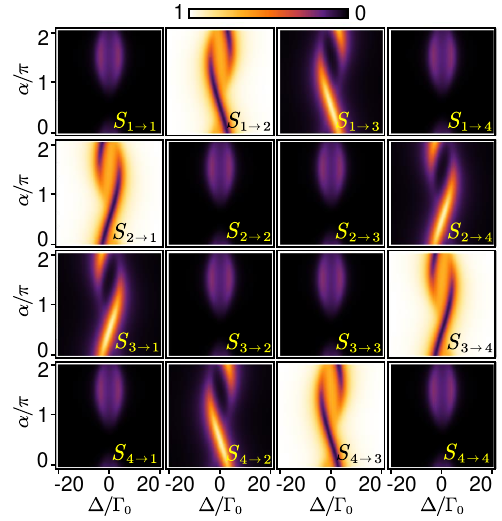}
    \caption{Schematic diagrams of the scattering matrix in a $\nabla$-type giant atom. The scattering probabilities for a four-port $\nabla$-type giant atom are depicted across 16 matrix elements, showing their variations with detuning $\Delta$ and phase shift $\alpha$. Each panel represents an element of the scattering matrix $S_{n\rightarrow m}$. Parameters include the interaction strength $\Omega = 8\Gamma_0$, decay rates $\gamma_{e1} = \gamma_{e2} = 0$, and propagation phases $\phi_{a, b} = \theta = \pi/2$.}
    \label{fig:S4x4}
\end{figure}

To describe the scattering behavior in a general way, the scattering amplitudes can be organized into a $4 \times 4$ scattering matrix as follows
\begin{equation}
|\psi_{\mathrm{ph}}\rangle^{\mathrm{out}} = \underset{\text{Induced by the Giant Atom}}{\underbrace{\left( \begin{matrix}
	s_{1\rightarrow 1} & s_{1\rightarrow 2} & s_{1\rightarrow 3} & s_{1\rightarrow 4} \\
	s_{2\rightarrow 1} & s_{2\rightarrow 2} & s_{2\rightarrow 3} & s_{2\rightarrow 4} \\
	s_{3\rightarrow 1} & s_{3\rightarrow 2} & s_{3\rightarrow 3} & s_{3\rightarrow 4} \\
	s_{4\rightarrow 1} & s_{4\rightarrow 2} & s_{4\rightarrow 3} & s_{4\rightarrow 4} \\
\end{matrix} \right)}} |\psi_{\mathrm{ph}}\rangle^{\mathrm{in}}.
\end{equation}

Such a matrix $s_{4\times4}$ can apply to the case of any four-port incident photon, and the corresponding outgoing photon state can be obtained. To simplify, we assume that the transition frequencies satisfy $\omega_{e_1} = \omega_{e_2} = \omega_e$. The scattering matrix can be conveniently illustrated by plotting the scattering probabilities $S_{i\rightarrow j}=\left| s_{i\rightarrow j} \right|^2$ versus parameters like the detuning $\Delta=\omega-\omega_e$ and the coupling phase $\alpha$, as shown in Fig.~\ref{fig:S4x4}. Furthermore, the phases accumulated by photons propagating between the two coupling points in waveguides read $\phi_{a,b}=(k_0+k_{\Delta})d_{a,b}={\phi_0}_{a,b}+\tau_{a,b}\Delta$, with ${\phi_0}_{a,b}=k_0d_{a,b}$ and $\tau_{a,b}=d_{a,b}/v_g$. Note that we here adopt the Markovian approximation, and the propagation time $\tau_{a,b}$ is sufficiently small to be neglected.


\section{Single photon scattering of the giant-atom node}
\label{Single photon scattering of giant quantum node}

So far, we have developed a four-port scattering model for the giant atom to study photon scattering. In this section, we examine the case where the giant atom is symmetrically coupled to the dual-rail waveguide, each having identical coupling strengths to the waveguide and coupling phases that match the propagation phases. Accordingly, we set the coupling strength $g_{1,2,3,4}=g_0$, and the phase $\theta_2=\theta_3,\theta_1=\theta_4,\phi_{a,b}=\theta=\theta_{2,3}-\theta_{1,4}$. The exact solution for each $s_{1\rightarrow n}$ port can be obtained
\begin{equation}
    \begin{aligned}
s_{1\rightarrow 1}=&\frac{-i\varGamma _0\left( e^{2i\theta}+\beta \right)}{\varDelta +i\frac{\gamma _{e_1}}{2}-\beta \Omega e^{i\alpha}+2i\Gamma _0\left( 1+\beta \right)},
\\
s_{1\rightarrow 2}=&\frac{-i\varGamma _0\left( 1+\beta \right)}{\varDelta +i\frac{\gamma _{e_1}}{2}-\beta \Omega e^{i\alpha}+2i\Gamma _0\left( 1+\beta \right)}+1,
\\
s_{1\rightarrow 3}=&\frac{-i\varGamma _0\left( 1+\beta \right) e^{i\theta}}{\varDelta +i\frac{\gamma _{e_1}}{2}-\beta \Omega e^{i\alpha}+2i\Gamma _0\left( 1+\beta \right)},
\\
s_{1\rightarrow 4}=&\frac{-i\varGamma _0\left( e^{i\theta}+\beta e^{-i\theta} \right)}{\varDelta +i\frac{\gamma _{e_1}}{2}-\beta \Omega e^{i\alpha}+2i\Gamma _0\left( 1+\beta \right)},
    \end{aligned}
\end{equation}
where $\Delta=E-\omega_{e}=\omega_0 + v_g k-\omega_{e}$ is the detuning between the incident photon and the atom.  $\Gamma_{0}=g_{0}^2/v_g$ describes the energy relaxation rate of excited states decaying into the waveguide. The relation between the two excitation amplitudes $u_{e_1}$ and $u_{e_2}$ in the giant atom can be characterized by the parameter $\beta$, which describes the ratio of $u_{e_2}$ to $u_{e_1}$ and indicates the occupation number imbalance between two excited states, i.e.,
\begin{equation}
u_{e_2}=\beta u_{e_1}=\frac{\beta g_0e^{-i\theta _2}}{\varDelta +i\frac{\gamma _{e_1}}{2}-\beta \Omega e^{i\alpha}+2i\Gamma _0\left( 1+\beta \right)},
\end{equation}
where
\begin{equation}
\label{beta-equation}
\beta =\frac{u_{e_2}}{u_{e_1}}=\frac{\varDelta +i\frac{\gamma _{e_1}}{2}+\Omega e^{-i\alpha}+2i\Gamma _0\left[ 1-e^{2i\theta} \right]}{\varDelta +i\frac{\gamma _{e_2}}{2}+\Omega e^{i\alpha}}.
\end{equation}
This result shows that in the case without external decay (i.e., $\gamma_{e_1}=\gamma_{e_2}=0$), the imaginary part in $\beta$ is governed by the traveling phase $\theta$ and the interatomic phase $\alpha$. It is worth mentioning that the term $2i\Gamma_0\left( 1-e^{2i\theta} \right)$ in Eq.~(\ref{beta-equation}) characterizes the inter-conversion of two excited states through the waveguide-induced interaction. 

\begin{figure}[tpb]
    \centering
    \includegraphics[width=0.96\linewidth]{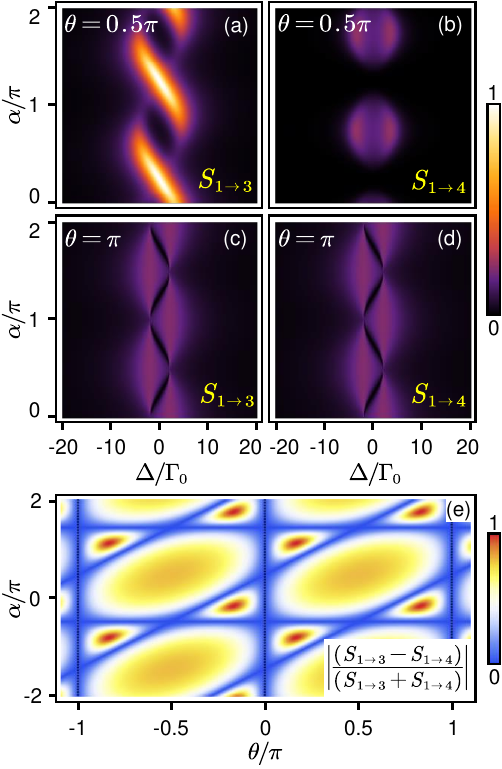}
    \caption{Unidirectional scattering behaviors in the upper waveguide of the $\nabla$-type giant-atom node. Scattering probabilities are shown for $\theta = 0.5\pi$ (a, b) and $\theta = \pi$ (c, d), alongside a heatmap of the unidirectional coefficient $U_b$ (e), where the dashed line marks $\theta = n\pi$.}
    \label{fig:S13S14}
\end{figure}

\begin{figure*}[tpb]
    \centering
    \includegraphics[width=1.0\linewidth]{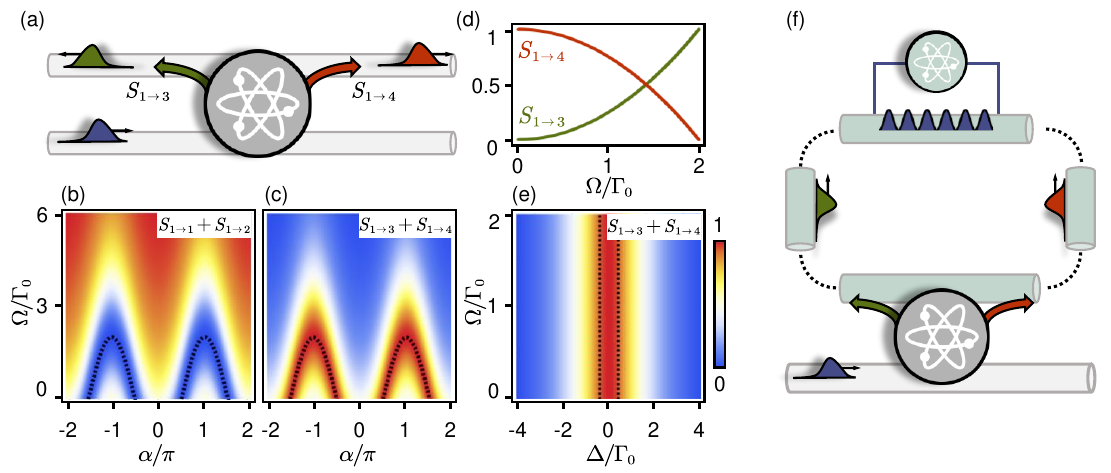}
    \caption{Trans-waveguide targeted routing. (a) Schematic diagram of the trans-waveguide targeted router, where photons are incident from port 1, routed to the upper waveguide, and propagate along the left (right) direction to exit from port 3 (4). Scattering amplitudes of the outgoing photon in the lower ($S_{1\rightarrow1}+S_{1\rightarrow2}$) (b) and upper ($S_{1\rightarrow3}+S_{1\rightarrow4}$) (c) waveguide versus interatomic phase $\alpha$ and coupling strength $\Omega$, where the black dashed curves show the phase-matching condition for the perfect trans-waveguide. (d) Scattering coefficient of leftward-propagating photons ($S_{1\rightarrow3}$) and rightward-propagating photons ($S_{1\rightarrow4}$) versus the interatomic coupling strength $\Omega$  under perfect trans-waveguide condition ($S_{1\rightarrow3}+S_{1\rightarrow4}=1$). (e) Scattering rate ($S_{1\rightarrow3}+S_{1\rightarrow4}$) to the upper waveguide versus incident photon detuning $\Delta$ and interatomic coupling strength $\Omega$ under perfect trans-waveguide routing parameters, where the black dashed lines indicate the scattering bandwidth ($\Delta\sim\Gamma$). (f) Application of trans-waveguide routing to realize the storage of a flying photon qubit. Other parameters are: atomic dissipation rates $\gamma_{e_1}=\gamma_{e_2}=0$, the ratio of two atomic emission rates into different waveguides $\Gamma_1/\Gamma_2=1$, and the accumulated phase of the photon traveling in the dual-rail waveguide $\phi_a=\phi_b=\pi/2$.}
    \label{fig:RoutingNew1314}
\end{figure*}

Figures~\ref{fig:S13S14}(a)-\ref{fig:S13S14}(d) show the scattering probabilities $S_{1\rightarrow 3}$ and $S_{1\rightarrow 4}$ with detuning $\Delta$ and interaction phase $\alpha$ for a one-port incident photon with modulated phase $\theta$, where the scattering probability can be calculated by $S_{i\rightarrow j}=\left| s_{i\rightarrow j} \right|^2$. 

When the coupling phase $\theta=\pi/2$, the term $2i\Gamma_0\left(1-e^{2i\theta}\right)$ becomes $4i\Gamma_0$, indicating the most substantial interatomic interaction caused by the waveguide. This interaction induces chirality of the photons propagating left and right in the upper waveguide $W_b$, as demonstrated by the distinct scattering rates $S_{1\rightarrow 3}$ and $S_{1\rightarrow 4}$ shown in Figs.~\ref{fig:S13S14}(a) and \ref{fig:S13S14}(b). 

Conversely, at $\theta=\pi$, the value of $2i\Gamma_0\left(1-e^{2i\theta}\right)$ reduces to 0, and the chirality of the left and right propagating photons within the upper waveguide $W_b$ disappears entirely, as shown in Figs.~\ref{fig:S13S14}(c) and \ref{fig:S13S14}(d). These results confirm that the interaction between the waveguide and the giant atom can induce output-photon chirality in the waveguide.

We define $U_b = \left|S_{1\rightarrow 3} - S_{1\rightarrow 4}/{S_{1\rightarrow 3} + S_{1\rightarrow 4}}\right|$ to quantify the directional transmission effectively in the upper waveguide. The graph of $U_b$, in relation to the detuning $\Delta$ and the phase $\theta$, as illustrated in Fig.~\ref{fig:S13S14}(e), suggests that a nontrivial scattering process is feasible with appropriate values of $\theta$ and $\phi_{a,b}$.


\section{Open system routing through the tunable giant-atom node}
\label{new routing chapter}

In this section, we aim to implement an ``open system" routing process using this giant atom in contrast to the ``coherent" routing via the stationary bus mode~\cite{zhouRealizingAlltoallCouplings2023}. Specifically, we analyze the routing scattering behavior by using a photon incident from port 1 (same in other cases). The routing process can be divided into two distinct regimes, namely the trans-waveguide regime and the unidirectional regime, either tailored to specific applications within quantum networks. Compared to existing approaches~\cite{yanTargetedPhotonicRouters2018,PhysRevA.105.023712,PhysRevResearch.2.043048} that rely on the chiral coupling between atoms and waveguides, which requires precise nanofabrication of photonic structures to maintain the necessary optical confinement and alignment, our interference-based method facilitates complete transmission without such challenging dependencies. 

\subsection{Trans-waveguide regime}
\begin{figure*}[tpb]
    \centering
    \includegraphics[width=1.0\linewidth]{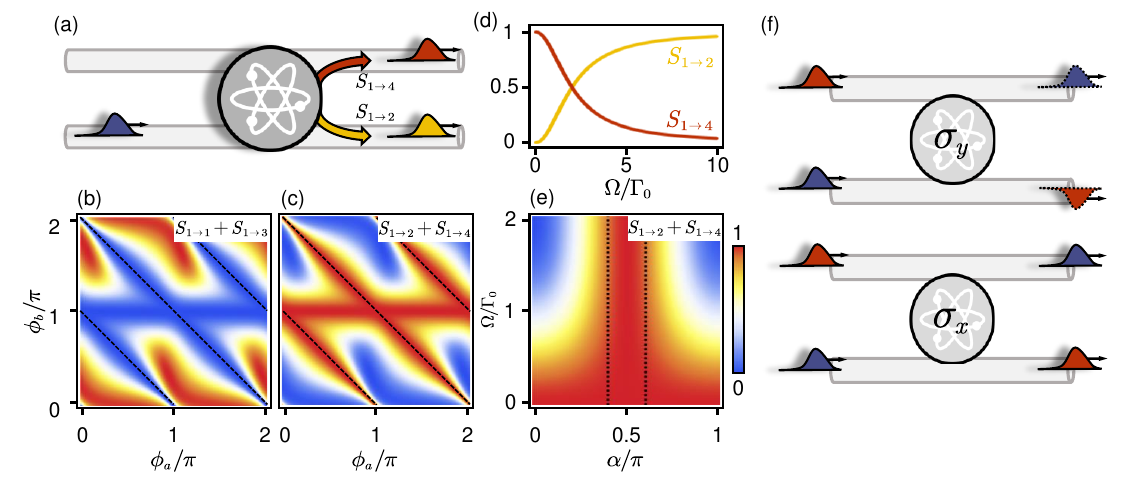}
    \caption{Unidirectional-waveguide targeted routing. (a) Schematic of unidirectional routing, where photons are incident from port 1, routed to the upper waveguide (lower waveguide), and propagate to the right exiting from port 2 (4). Scattering intensity versus photon propagation phase $\phi_a$ and $\phi_b$ for outgoing photons propagating to the left ($S_{1\rightarrow1}+S_{1\rightarrow3}$) (b) and the right ($S_{1\rightarrow2}+S_{1\rightarrow4}$) (c) after scattering. (d) Scattering coefficient versus coupling strength $\Omega$ for rightward-propagating photons in the upper waveguide  ($S_{1\rightarrow4}$) and the lower waveguide ($S_{1\rightarrow2}$) under perfect unidirectional conditions. (e) Shows the scattering intensity versus propagation phase mismatch $\delta \phi$ and interatomic coupling strength $\Omega$ for outgoing photons to the right. (f) Schematic of a quantum gate constructed by a giant atom, which shows the photon scattering process of the gates $\sigma_x$ and $\sigma_y$. Other parameters are: atomic dissipation rates $\gamma_{e_1}=\gamma_{e_2}=0$, the ratio of two atomic emission rates into different waveguides $\Gamma_1/\Gamma_2=1$, and the accumulated phase of the photon traveling in dual-rail waveguide $\phi_a=\phi_b=\pi/2$.}
    \label{fig:UnidirectionalRouting1214}
\end{figure*}
In quantum networks, parallel channels can contribute to improving system connectivity and communication efficiency~\cite{burgarth2005conclusive}. To construct parallel channels in the waveguide system, it is necessary to transfer flying photons from one waveguide to the other, which can be achieved in our model. To achieve perfect trans-waveguide transmission, the conditions $s_{1\rightarrow 1} = s_{1\rightarrow 2} = 0$ must be met, as illustrated in Fig.~\ref{fig:RoutingNew1314}(a), ensuring that photons incident from port 1 do not exit at ports 1 and 2 of the original waveguide. Under these conditions, the phase and strength can be derived through system parameters. The propagating phases and coupling phases for the lower and upper waveguides should satisfy $\phi_{a}=\phi_{b}=\pi/2$, $\theta_{1}=-\theta_{2}$, and $\theta_{4}-\theta_{3} = 2\alpha +\theta_{2}-\theta_{1}$, respectively. The constraint for the coupling strength becomes
\begin{equation}
\Omega = 2\cos(2\theta_{2}+\alpha)\Gamma_{12}, 
\end{equation}
where $\Gamma_{12}=g_{1}g_{2}/{v_{g}}$ represents the collective energy relaxation rate of the giant atom into the output channels. 
In such a parameter space, setting $\theta_{3}=\theta_{2}=\pi$ gives two degrees of freedom, namely $\alpha$ and $\Omega$. 

Figures~\ref{fig:RoutingNew1314}(b) and \ref{fig:RoutingNew1314}(c) illustrate the dependence on $\alpha$ and $\Omega$ of the scattering probabilities $S_{1\rightarrow 1}+S_{1\rightarrow 2}$ and $S_{1\rightarrow 3}+S_{1\rightarrow 4}$. 
These confirm the constraints for perfect trans-waveguide transmission, i.e., $S_{1\rightarrow 3}+S_{1\rightarrow 4}=1$. Furthermore, we calculate the resilience to $\Delta$ and $\Omega$ of the efficiency $S_{1\rightarrow 3}+S_{1\rightarrow 4}$, as depicted in Fig.~\ref{fig:RoutingNew1314}(e), revealing a favorable conversion efficiency within the $\Delta\sim\Gamma$ range.

In the detuning-free case, the giant-atom configuration can function as a tunable beamsplitter by adjusting the interatomic coupling strength $\Omega$, as shown in Fig.~\ref{fig:RoutingNew1314}(d), 
which allows us to modulate the one-way emission of photons in the upper waveguide. This unidirectional photon transport has been extensively studied in a field known as ``chiral quantum optics"~\cite{lodahl2017chiralQO}. It has been achieved in artificially designed structures, where light interacts with different atomic energy levels to enable directional coupling between photons and emitters~\cite{bliokh2015spinorbitlight}.  
Unlike previous works, our approach enables unidirectional photon emission in the upper waveguide by modulating the giant-atom node, which offers a possibility for targeted transmission.


Additionally, by selecting an appropriate coupling strength ($\Omega=1.4\Gamma$), our approach enables the giant atom to function as a 50:50 beamsplitter, which emits a superposition of left and right propagating photon states at the upper waveguide. Based on this, we can integrate the system with the configurations proposed in Ref.~\cite{2024guogiantatomcatchphoton}, where it is demonstrated that such superposed states can be effectively trapped within a specially designed giant atom at the far end of the waveguide, as shown in Fig.~\ref{fig:RoutingNew1314}(f). This design facilitates this integration, thereby enabling the construction of a robust `quantum storage area' for a flying qubit. Such quantum storage can function as a register in a quantum network, providing a crucial element for scalable and efficient quantum communication and computation, thus expanding the applicability of our system in broader scenarios.

\subsection{Unidirectional-waveguide regime}\label{unisec}
In the dual-rail waveguide system, in addition to trans-waveguide routing, there is also a unidirectional routing. This routing is designed to enable one-way photon propagation within the dual-rail setup and control the scattering ratio of photons in the upper and lower waveguides, as illustrated in Fig.~\ref{fig:UnidirectionalRouting1214}(a).

The key to rightward unidirectional routing is the anti-reflective condition for photon transmission in the dual-rail waveguide system, denoted by $s_{1\rightarrow 1} = s_{1\rightarrow 3} = s_{3\rightarrow 1} = s_{3\rightarrow 3} = 0$. Under the above condition, we identify the phase-matching constraints for the giant atom to satisfy
\begin{equation}
\label{phasematchuni}
    \begin{aligned}
    k \pi &= \phi_{a} + \phi_{b} , \\
    k \pi &= \theta_{1} - \theta_{2} + \theta_{3} - \theta_{4}, \\
    \alpha &= \theta_{4} - \theta_{3}.
    \end{aligned}
\end{equation}
To verify the above results, the rightward and leftward scattering probabilities with respect to $\phi_{a,b}$ are calculated under the constraints $ \theta_{2} + \theta_{4}+k \pi= \theta_{1} + \theta_{3}$ and $ \alpha= \theta_{4} - \theta_{3}$, as shown in Figs.~\ref{fig:UnidirectionalRouting1214}(b) and \ref{fig:UnidirectionalRouting1214}(c). Perfect unidirectional transmission can be realized at the black dashed line in the figure, which is consistent with the constraint $\phi_{a}+\phi_{b}=k \pi$.

Furthermore, we also examined the performance of this system by calculating its efficiency versus $\alpha$ and $ \Omega$, as shown in Fig.~\ref{fig:UnidirectionalRouting1214}(e). This reveals that, despite potential imperfections during the system preparation, our method consistently maintains high fidelity, particularly when $|\delta \alpha | \leqslant  0.1 \pi$.

To simplify the experimental setup and improve its feasibility, we set the propagating phase at the coupling points to $\phi_{a} = \phi_{b} = \pi/2$. This condition allows us to derive precise constraints for the coupling phase of both waveguides. Therefore, for the lower and upper waveguides, the conditions become $\theta_{2} - \theta_{1} = -\pi/2 - \alpha$ and $\theta_{4} - \theta_{3} = -\pi/2 + \alpha$, respectively.

Now, we calculate the scattering probabilities $S_{12}$ and $S_{14}$ versus coupling strength $\Omega$ in the absence of detuning. Figure~\ref{fig:UnidirectionalRouting1214}(d) illustrates this relation and demonstrates that the modulation of $\Omega$ allows for precise control over the routing of photons entering at port 1 to either port 2 or port 4. This property benefits targeted emission control between the upper and lower ports by adjusting the coupling strength, which acts as a unidirectional beamsplitter. 

The unidirectional regime enables the path encoding method~\cite{levy2024passive}, where photons in the upper waveguide $W_b$ can be encoded as $|\mathrm{up}\rangle$, and the lower waveguide $W_a$ as $|\mathrm{down}\rangle$, as shown in Fig.~\ref{fig:UnidirectionalRouting1214}(f). Each scattering of a photon by the giant atom corresponds to a unitary operation on the path-encoded photon qubit.
Substituting Eq.~(\ref{phasematchuni}) into the initial scattering matrix yields the expressions for $s_{1\rightarrow 2},s_{1\rightarrow 4},s_{3\rightarrow 2},s_{3\rightarrow 4}$:
\begin{equation}
    \begin{aligned}
    s_{1\rightarrow 2} &= s_{3\rightarrow 4} = \frac{\Omega}{2i \Gamma_{12} + \Omega}, \\
    s_{1\rightarrow 4} &= \frac{2i\Gamma_{12}}{2i\Gamma_{12} + \Omega} \exp\{i(\alpha - \theta_{1} + \theta_{3})\}, \\
    s_{3\rightarrow 2} &= \frac{2i\Gamma_{12}}{2i\Gamma_{12} + \Omega} \exp\{-i(\alpha - \theta_{1} + \theta_{3})\}.
    \end{aligned}
\end{equation}
The giant-atom node here corresponds to a unitary evolution of the path-encoded photon qubit $U(\Omega,\theta)$,
\begin{equation}
    U(\Omega,\theta)=
    \left(
    \begin{array}{cc}
    s_{3\rightarrow 4}  & s_{1\rightarrow 4} \\
        s_{3\rightarrow 2} & s_{1\rightarrow 2}
    \end{array}
    \right).
\end{equation}
Further, the above operation can be written in the following form
\begin{equation}
\begin{aligned}
	U(\Omega ,\theta )&=U(\delta (\Omega ),\phi (\theta ))\\
	&=e^{-i\delta}\left( \begin{matrix}
	\cos\delta&		ie^{i\phi}\sin\delta \\
	ie^{-i\phi}\sin\delta &		\cos\delta\\
\end{matrix} \right) ,\\
\end{aligned}
\end{equation}
where $\delta = \arctan(2\Gamma_{12}/\Omega)$ and $\phi = \alpha - \theta_1 + \theta_3$.

Various quantum gates can be constructed by adjusting parameters in the giant-atom node. For example, setting $\delta=0$ results in the identity operation $U=1$, preserving the initial photon states. Choosing $\delta=\pi/2$ and $\phi=0$ yields $U=\sigma_x$, while $\delta=\pi/2$ and $\phi=-\pi/2$ gives a $\sigma_y$ gate. 

Figure~\ref{fig:UnidirectionalRouting1214}(f) illustrates the implementation of $\sigma_x$ and $\sigma_y$ gates using a giant-atom node in a dual-rail waveguide. An arbitrary unitary gate can be achieved by sequentially applying two giant-atom scatterings to a photon. For instance, a $\sigma_z$ gate can be constructed using two cascaded giant-atom nodes, i.e., $U_1 U_2 = i \sigma_z$, with parameters $\delta_1=\pi/2$, $\phi_1=0$ and $\delta_2=\pi/2$, $\phi_2=-\pi/2$ for the first and second nodes, respectively.

Furthermore, the interatomic interactions in the giant atom configuration can be controlled by an auxiliary qubit~\cite{guimond2020unidirectional}. The state of the auxiliary qubit influences the effective coupling strength $\Omega$, enabling a controlled quantum gate between the qubit and the scattered path-encoded photon qubit, facilitating photon-qubit entanglement. Suppose the auxiliary qubit is coupled to the system. The effective coupling strength $\Omega$ becomes zero when the auxiliary qubit is in state $|1\rangle$ (i.e., $\delta=\pi/2$), and increases to $10 \Gamma_{12}$ when the auxiliary qubit is in state $|0\rangle$ (i.e., $\delta=0$). Under these conditions, the giant atom configuration functions as a controlled-NOT (CNOT) gate between the auxiliary qubit and a photon qubit.

\subsection{Quantum teleportation between two giant-atom nodes}

\begin{figure}[tpb]
    \centering
    \includegraphics[width=1.0\linewidth]{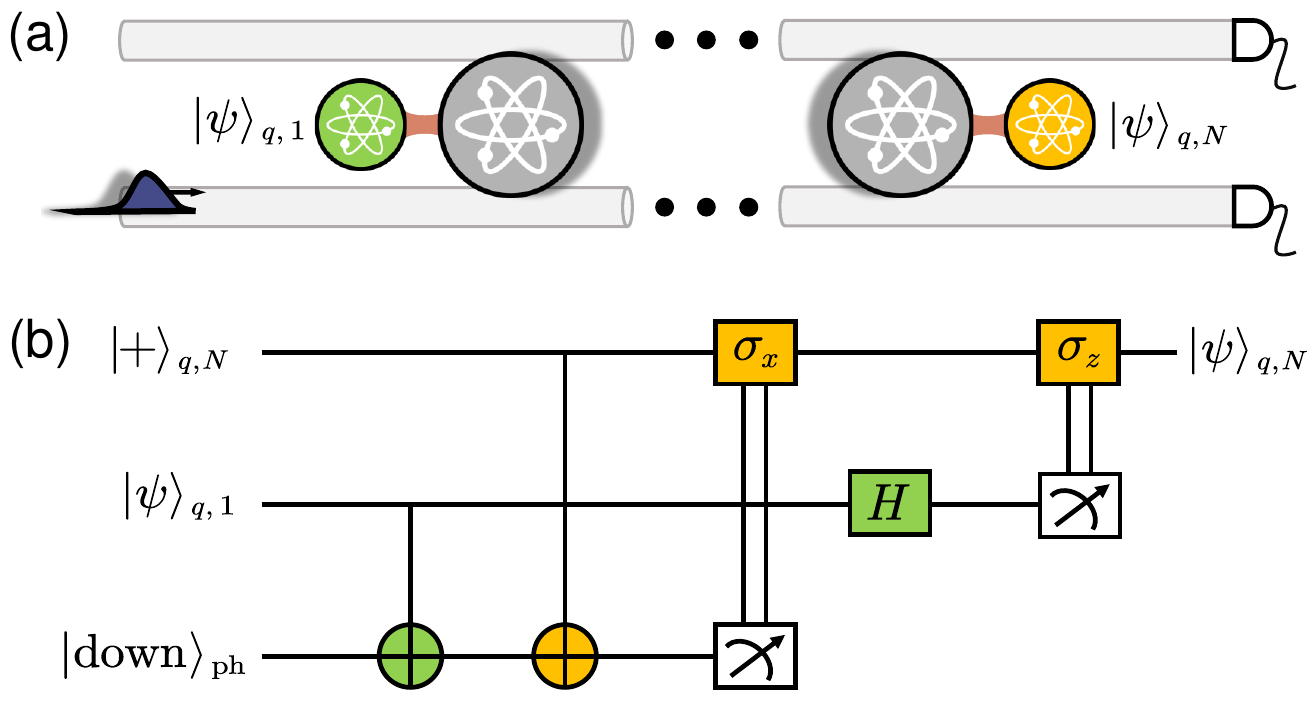}
    \caption{Quantum teleportation using giant atoms. (a) Setup of a quantum network consisting of giant-atom nodes and a dual-rail waveguide. (b) Schematic quantum circuit diagram of quantum state transfer from giant-atom node 1 to node \textit{N}, where the green part and the orange part indicate the operation on giant atoms 1 and \textit{N}, respectively. By detecting photons in the waveguide and reading the initial state $|\psi\rangle_{q,1}$ of the giant-atom node 1, the quantum state is finally transferred to the quantum state $|\psi\rangle_{q,N}$ at the output. This protocol leverages the concept of quantum teleportation, allowing the quantum state transfer between two parties through entangled Bell states.} 
    \label{fig:quantumnetwork}
\end{figure}

Inspired by quantum teleportation, we propose a theoretical approach that utilizes unidirectional giant-atom nodes for quantum state transfer. In our approach, quantum information is path-encoded into flying photons~\cite{2015scienceUniversallinearoptics} and directed through these nodes. By leveraging the unidirectional arbitrary quantum gates described earlier, the nodes facilitate the transfer of quantum states. This protocol draws from the principles of quantum teleportation~\cite{huProgressQuantumTeleportation2023, bouwmeesterExperimentalQuantumTeleportation1997}, enabling the transfer of a quantum state between two entities through shared entangled Bell states.

In this approach, entanglement between giant-atom node 1 and a photon is established through a CNOT operation from the auxiliary qubit of giant-atom node 1 to the photon. This is followed by generating three-body entanglement via another CNOT operation from the auxiliary qubit of giant-atom node $N$ to the previously entangled photon. The process concludes with the measurement of the photon, entangling the two auxiliary qubits in giant-atom nodes 1 and $N$, enabling quantum teleportation between them. The process concludes with the measurement of the photon, which entangles the two auxiliary qubits in giant-atom nodes 1 and $N$ and enables quantum teleportation between them.

As discussed in Sec.~\ref{unisec}, dual-waveguide giant atoms impact photons by acting as either $\sigma_x$ or $\mathbf{1}$ gates when the auxiliary qubit is in different quantum states, leading to the auxiliary qubits' CNOT gate on photon qubits. Specifically, when the input state of the auxiliary qubit is $|\psi\rangle_{q} = a_{0}|0\rangle + a_{1}|1\rangle$ and the photon is in the state $|\psi\rangle_{\mathrm{ph}} = b_{0}|\mathrm{down}\rangle + b_{1}|\mathrm{up}\rangle$, the resulting state is $|\psi\rangle = a_{0}|0\rangle \otimes (b_{0}|\mathrm{down}\rangle + b_{1}|\mathrm{up}\rangle) + a_{1}|1\rangle \otimes (b_{0}|\mathrm{up}\rangle + b_{1}|\mathrm{down}\rangle)$. By combining these controlled quantum gates with the modulation of auxiliary qubits (H-gate, $\sigma_{x}$-gate, $\sigma_{z}$-gate), a quantum state transfer network, as depicted in Fig.~\ref{fig:quantumnetwork}(b), can be implemented. 

This method enables the remote transfer of the quantum state $|\psi\rangle_{q1} = a_{0}|0\rangle + a_{1}|1\rangle$ from a qubit within giant atom 1 to another in giant atom \textit{N}, as depicted in Fig.~\ref{fig:quantumnetwork}(a). This process is executed through a quantum circuit, illustrated in Fig.~\ref{fig:quantumnetwork}(b), which begins with two controlled operations executed via photon scattering involving giant-atom nodes 1 and \textit{N}.
Subsequent steps include single-qubit measurements and manipulations on photons and nodes, enabled by photon detectors and control lines connected to the giant atoms. Conditional $\sigma_{x}$ and $\sigma_{z}$ operations are performed on node \textit{N} depending on whether the photon or the auxiliary qubit in node 1 is in the $|\mathrm{up}\rangle_{\mathrm{ph}}$ or $|1\rangle_{q}$ states, respectively. 

This approach demonstrates the capability of giant-atom nodes to perform quantum teleportation, highlighting their potential applications in quantum information transfer and distribution.


\section{A Circulator through Engineered Nonreciprocity}
\label{photon nonreciprocal scattering}

\subsection{Nonreciprocal scattering using a giant atom}

\begin{figure}[tpb]
    \centering
    \includegraphics[width=0.96\linewidth]{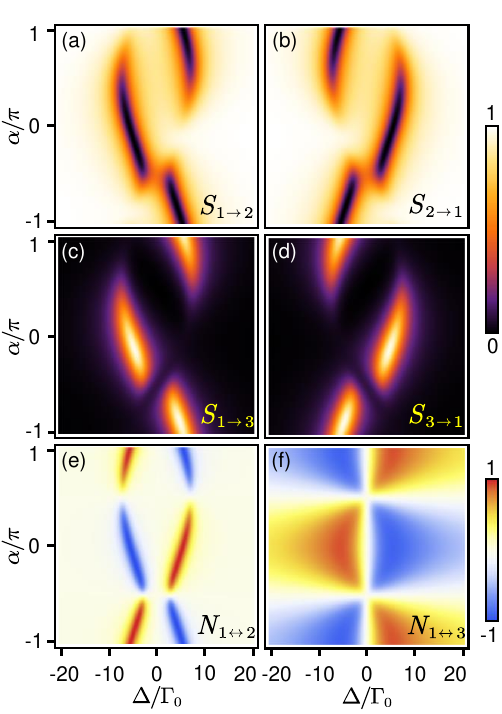}
    \caption{Nonreciprocal scattering behaviors of a $\nabla$-type giant-atom node. Scattering probabilities: (a) $S_{1\rightarrow 2}$; (b) $S_{2\rightarrow 1}$; (c) $S_{1\rightarrow 3}$; (d) $S_{3\rightarrow 1}$. Nonreciprocal coefficients: (e) $N_{1\leftrightarrow 2}$; (f) $N_{1\leftrightarrow 3}$ as functions of the detuning $\Delta$ and interaction phase $\alpha$, with interaction strength $\Omega=5\Gamma_0$.}
    \label{fig:nonreciprocity}
\end{figure}

Nonreciprocal devices, which exhibit inherent asymmetry between their forward and backward propagation directions, are currently of broad interest for enabling new applications in acoustics~\cite{SoundIsolation}, photonics~\cite{2016ScienceQuantumopticalcirculator,doi:10.1126/science.aaa9519}, and superconducting circuits~\cite{PhysRevLett.130.037001}. In these examples, they are essential for isolating qubits from noisy channels and serve as key components in photon circulators within quantum circuits. In this section, we investigate the phenomena of nonreciprocal scattering induced by giant-atom nodes and their application in a quantum circulator. 

\begin{figure*}[tpb]
    \centering
    \includegraphics[width=0.8\linewidth]{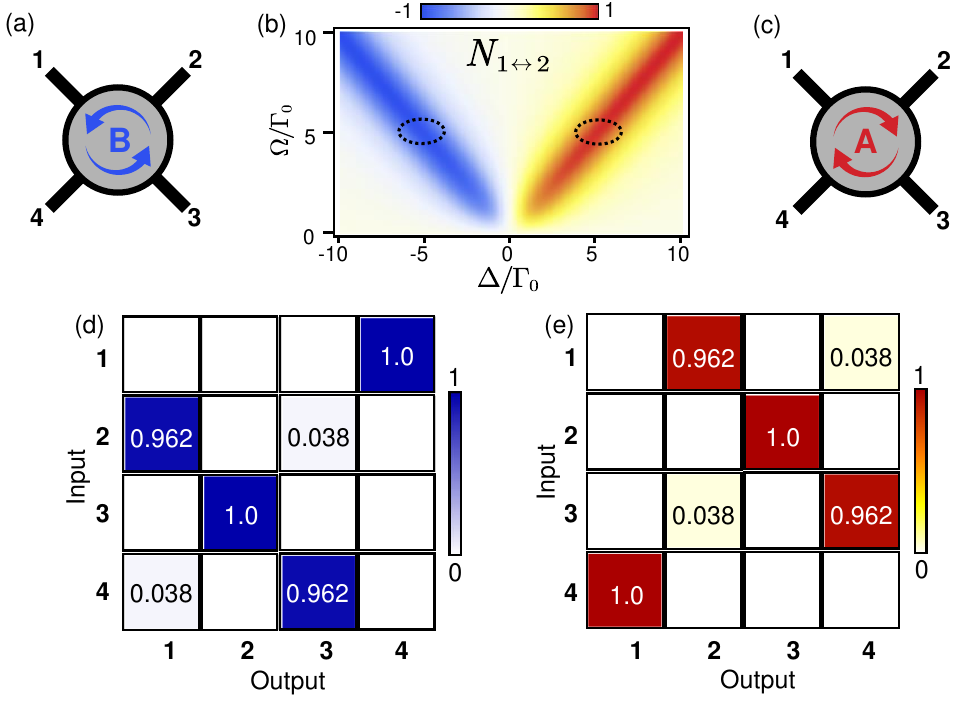}
    \caption{Giant-atom node circulator. The circulator has two modes: counterclockwise (a) and clockwise (c). (b) Nonreciprocal coefficient $N_{i\leftrightarrow j}$ versus the detuning $\Delta$ and interatomic coupling strength $\Omega$. The two different circulating modes correspond to different regions on the photon scattering nonreciprocity map, where the blue region corresponds to the counterclockwise mode (a) and the red region corresponds to the clockwise mode (c). (d) and (e) are the scattering matrices $S_{nm}$ under the conditions $\Omega = 5\Gamma_0$, $\Delta = \mp 5\Gamma_0$, where the red color block (d) and the blue color block (e) correspond to the different circulating modes. The form of these scattering matrices are consistent with the mathematical form of the circulator defined in Eq.~\eqref{circulator-matrices}.}
    \label{fig:circulator}
\end{figure*}
For simplicity, we set the phase-matching condition of different scattering paths by choosing $\phi_{a} = \phi_{b} = \theta = \theta_1 - \theta_2 = \theta_4 - \theta_3 = \pi/2$. Applying scattering theory and black-box quantization methods from Section~\ref{model and methods}, we determine the scattering probabilities $S_{2\rightarrow 1}$ and $S_{3\rightarrow 1}$ for photons incident from ports 2 and 3, respectively. As shown in Figs.~\ref{fig:nonreciprocity}(a)--\ref{fig:nonreciprocity}(d), we plot the scattering probabilities as functions of the detuning $\Delta$ and interaction phase $\alpha$. To quantitatively analyze the nonreciprocity of the system, we define the nonreciprocal coefficient between port $i$ and port $j$ as:
\begin{equation}
    \label{nonreciprocity-coefficient}
    N_{i\leftrightarrow j} = \frac{S_{i\rightarrow j} - S_{j\rightarrow i}}{S_{i\rightarrow j} + S_{j\rightarrow i}}, \quad i \ne j.
\end{equation}
Figures~\ref{fig:nonreciprocity}(e) and \ref{fig:nonreciprocity}(f) show that the giant-atom node exhibits tunable nonreciprocity coefficients in the nonzero detuning region.
Due to their tunable nonreciprocity, giant atoms are beneficial for constructing nonreciprocal devices.

\subsection{Realizing a circulator with a giant atom}

A circulator can be defined by the following scattering matrices~\cite{jalasWhatWhatNot2013}:
\begin{equation}
\label{circulator-matrices}
    S_{\text{ccw}} = \begin{pmatrix}
        0 & 0 & 0 & 1 \\
        1 & 0 & 0 & 0 \\
        0 & 1 & 0 & 0 \\
        0 & 0 & 1 & 0 \\
    \end{pmatrix}, \quad
    S_{\text{cw}} = \begin{pmatrix}
        0 & 1 & 0 & 0 \\
        0 & 0 & 1 & 0 \\
        0 & 0 & 0 & 1 \\
        1 & 0 & 0 & 0 \\
    \end{pmatrix},
\end{equation}
representing counterclockwise and clockwise circulators, respectively.

To realize a circulator using our giant-atom node, we search for the condition under which $S_{1\rightarrow 1} = S_{2\rightarrow 2} = S_{3\rightarrow 3} = S_{4\rightarrow 4} = 0$, ensuring that there is no reflection at any port. Under the conditions ($\phi_a = \phi_b = \pi/2$) studied previously, we find that the phase parameters must satisfy the phase-matching conditions:
\begin{equation}
\begin{aligned}
\theta _2-\theta _1=-\pi/2 -\alpha, 
\\
\theta _4-\theta _3=-\pi/2 +\alpha.                
\end{aligned}
\end{equation}
By implementing these circulator phase conditions, we compute the nonreciprocal coefficients as functions of both detuning $\Delta$ and coupling strength $\Omega$, as illustrated in Fig.~\ref{fig:circulator}(b). This figure shows two modes of circulators corresponding to different circulating directions.

Without loss of generality, we choose the phases introduced by the waveguides as $\theta_1 = \theta_3 = 0$ and set $\alpha = \pi/2$. In the parameter interval where $\Delta = -\Omega$, corresponding to circulator mode 1, the photon circulation is as follows: photons entering at port 1 exit from port 4, photons entering at port 4 exit from port 3, photons entering at port 3 exit from port 2, and photons entering at port 2 exit from port 1. This circulation mode is represented by the blue region in Fig.~\ref{fig:circulator}(b) and is illustrated as a counterclockwise circulator in Figs.~\ref{fig:circulator}(a) and \ref{fig:circulator}(e).

In the interval where $\Delta = \Omega$, defining circulator mode 2, the photon flow is reversed: photons entering at port 1 exit from port 2, photons entering at port 2 exit from port 3, photons entering at port 3 exit from port 4, and photons entering at port 4 exit from port 1. This directional flow is indicated by the red region in Fig.~\ref{fig:circulator}(b) and is depicted as a clockwise circulator in Figs.~\ref{fig:circulator}(c) and \ref{fig:circulator}(d).

Therefore, we have utilized giant-atom nodes to implement directionally adjustable four-port circulators. Nonreciprocity is achieved through coherent phase cancellation among multiple paths within the giant-atom node, rather than by relying on externally introduced active devices. This mechanism does not require external strong driving, resulting in a compact and integrable design. Consequently, this advancement has the potential to enhance the development of large-scale, integrable on-chip quantum interconnections.


\section{Feasibility and Imperfections}
\label{sec:feasibility}

The giant-atom model has already been studied on multiple experimental platforms~\cite{gustafsson2014propagating,kannanWaveguideQuantumElectrodynamics2020,2023PhysRevXResonanceFluorescenceofaChiralArtificialAtom,anderssonNonexponentialDecayGiant2019}, among which superconducting circuits stand out as a promising candidate. Implementing the giant-atom model within this well-established platform offers a practical route for realizing our proposed system. In our approach, the three-level $\nabla$-type giant atoms can be constructed using identical qubits, which facilitate independent control of the two excited states. The gauge coupling between qubits and the waveguide, as well as between different qubits, is achieved through time-modulated magnetic flux techniques in inductive coupling~\cite{roushan2017chiral,cao2024parametrically,Wang_2022,2023PhysRevXResonanceFluorescenceofaChiralArtificialAtom,PhysRevApplied.6.064007}, allowing for precise control of interaction strength and phase. Detailed experimental implementations, computational models, and technical analysis are provided in Appendix~\ref{appendix:realization}.

However, in practical experiments, systems inevitably exhibit dissipation and various mismatches, leading to deviations from the ideal behavior. We investigate how imperfections, such as qubit decay $\gamma_{e}$, coupling strength mismatch $\delta_\Omega$, coupling phase mismatch $\delta_\alpha$, and propagating phase mismatch $\delta_\phi$, affect the fidelity of routing, circulators, and quantum gates. Here, the qubit decay $\gamma_{e}$ represents the rate at which the atom loses energy to the environment, the coupling strength $\Omega$ deviates from the optimal parameter $\Omega_{0}$ by a detuning $\delta_\Omega$, such that $\Omega = \Omega_{0} + \delta_\Omega$, the coupling phase mismatch $\delta_\alpha$ accounts for deviations in the phase of the coupling between qubits, and the propagating phase mismatch $\delta_\phi$ represents deviations in the phase during the propagation of photon.

\subsection{Routing Fidelity}

Initially, we calculate the fidelity of the router~\cite{yanTargetedPhotonicRouters2018}. Within the frameworks of unidirectional and trans-waveguide transmission regimes, the fidelities are defined as follows:
\begin{equation}
    \mathcal{F}_{\text{dire}} = \frac{S_{1\rightarrow 2} + S_{1\rightarrow 4}}{\sum_{i=1}^4 S_{1\rightarrow i}}, \quad
    \mathcal{F}_{\text{trans}} = \frac{S_{1\rightarrow 3} + S_{1\rightarrow 4}}{\sum_{i=1}^4 S_{1\rightarrow i}},
\end{equation}
where $S_{1\rightarrow j}$ denotes the scattering probability from port 1 to port $j$. The fidelities $\mathcal{F}_{\text{dire}}$ and $\mathcal{F}_{\text{trans}}$ are the ratios of photons correctly routed in the unidirectional and trans-waveguide transmission scenarios, respectively.

\begin{figure}[tpb]
    \centering
    \includegraphics[width=1.025\linewidth]{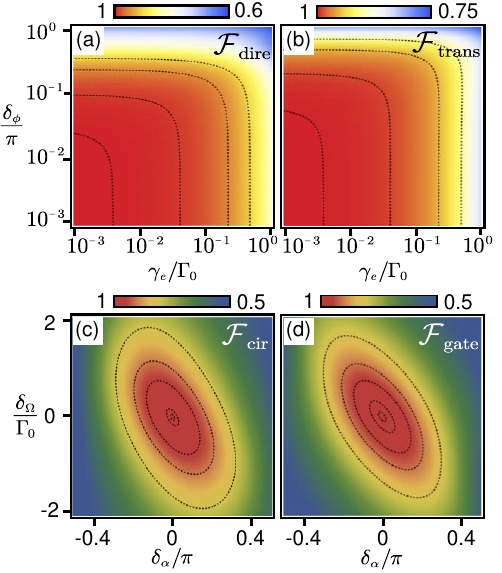}
    \caption{Fidelity influenced by decay rate and coupling mismatches in quantum operations. (a) Fidelity of unidirectional transmission and as a function of qubit decay rate $\gamma_e$ and propagating phase mismatch $\delta_\phi$. (b) Fidelity of trans-waveguide transmission across a waveguide with varying $\gamma_e$ and $\delta_\phi$. (c) Fidelity of a circulator depending on coupling strength mismatch $\delta_\Omega$ and coupling phase mismatch $\delta_\alpha$. (d) Fidelity of the $\sigma_{x}$ gate with respect to $\delta_\Omega$ and $\delta_\alpha$.}
    \label{fig:fidelity}
\end{figure}

Figures~\ref{fig:fidelity}(a) and~\ref{fig:fidelity}(b) depict the routing fidelities as functions of qubit decay $\gamma_{e}$ and propagating phase mismatch $\delta_\phi$, showing that fidelities are relatively insensitive to $\gamma_{e}$. Moreover, for both routing scenarios within the intervals $\delta_\phi/\pi \lesssim 0.1$ and $\gamma_{e}/\Gamma \lesssim 0.1$, the fidelity exceeds 0.99, indicating robust performance against imperfections.
\subsection{Circulator Fidelity}

The fidelity of a circulator is defined based on the trace distance between the actual scattering matrix $\tilde{T}$ and the ideal circulator scattering matrix $T^{\text{id}}$~\cite{2016ScienceQuantumopticalcirculator}:
\begin{equation}
    \mathcal{F}_{\text{cir}} = \frac{\mathrm{Tr}[\tilde{T} \cdot (T^{\text{id}})^{\mathrm{T}}]}{\mathrm{Tr}[T^{\text{id}} \cdot (T^{\text{id}})^{\mathrm{T}}]}.
\end{equation}
We take the clockwise circulator as an example to calculate the fidelity. Figure~\ref{fig:fidelity}(c) shows the circulator fidelity as a function of $\delta_\Omega$ and $\delta_\alpha$. The fidelity remains above 0.99 within the parameter ranges $\delta_\Omega/\Gamma \lesssim 0.1$ and $\delta_\alpha/\pi \lesssim 0.1$, demonstrating that the circulator performance is also resilient to these imperfections.

\subsection{Gate Fidelity}

Finally, we examine the fidelity of the $\sigma_{x}$ gate~\cite{NIELSEN2002249}, which represents a path-switching operation facilitated by the giant-atom node on photons. The average gate fidelity is defined as the average fidelity over various input states within the Hilbert space:
\begin{equation}
    \mathcal{F}_{\text{gate}} = \frac{1}{N} \sum_{i=1}^{N} |\langle \psi_{i} | V^{\dagger} U | \psi_{i} \rangle |^{2},
\end{equation}
where $V$ is the ideal gate operator, $U$ is the actual operation performed, and $|\psi_{i}\rangle$ are the input states. The resulting fidelity, shown in Fig.~\ref{fig:fidelity}(d), indicates that the gate operation maintains high fidelity in the same parameter ranges.

In summary, this result demonstrates that the proposed giant-atom system exhibits high fidelity in routing, circulator, and gate operations when subject to realistic imperfections such as qubit decay, coupling mismatch, and propagating mismatch. The robustness of the system within acceptable parameter ranges highlights its feasibility for practical implementations in superconducting quantum circuits. This resilience to imperfections is a promising step toward the development of scalable, high-fidelity quantum information processing devices.


\section{Conclusion} 
\label{conclusion}

We designed a universal giant atom routing node within a dual-rail waveguide, enabling on-demand photon transport between different ports. Achieving unidirectional information transfer is fundamental for constructing practical networks, and our proposed nodes function as circulators with tunable clockwise and counter-clockwise operations under varying parameter conditions. The nonreciprocity in our design is achieved through coherent phase cancellation rather than relying on externally introduced active devices as traditional circulators, eliminating the need for external strong driving and resulting in a compact and integrable structure. This simplification facilitates the integration of different modules using such nodes, paving the way for all-to-all quantum networks. Furthermore, by encoding information in the photon paths, we have demonstrated the potential of the proposed giant-atom nodes for long-distance quantum state transfer using flying photons. This application underscores the promising potential of giant atom systems as quantum nodes, paving the way for future applications in more complex and large quantum networks.


\section{Acknowledgments}

We thank Xin Wang for stimulating discussions. This work is supported by the the National Natural Science Foundation of China (Grant No. 12375025 and No. 11874432), the National Key R\&D Program of China (Grant No. 2019YFA0308200). F.N. is supported in part by: Nippon Telegraph and Telephone Corporation (NTT) Research, the Japan Science and Technology Agency (JST) [via the CREST Quantum Frontiers program Grant No. JPMJCR24I2, the Quantum Leap Flagship Program (Q-LEAP), and the Moonshot R\&D Grant Number JPMJMS2061], and the Office of Naval Research (ONR) Global (via Grant No. N62909-23-1-2074).

\appendix
\section{Solution of the scattering matrix in a four-port model of $\nabla$-type giant atoms}
\label{appendix:Solution of the scattering matrix}
\begin{figure}[htp]
    \centering
    \includegraphics[width=1\linewidth]{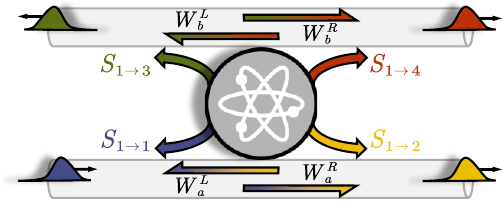}
    \caption{Schematic diagram of a photon incident into port 1 of the giant atom.}
    \label{fig:scatterdetail}
\end{figure}

Solving the eigenvalue equation $H|\Psi\rangle=E|\Psi\rangle$ yields the following equations
\begin{equation}
\label{con:dynamics equation}
\begin{aligned}
  E\Phi _{al}(x)=&\left( \omega_0+if_lv_g\frac{\partial}{\partial x} \right) \Phi _{al}(x)
\\
&+g_1\delta (x)e^{i\theta _1}u_{e2}+g_2\delta \left( x-d_a \right) e^{i\theta _2}u_{e1},
\\
E\Phi _{bl}(x)=&\left( \omega_0+if_lv_g\frac{\partial}{\partial x} \right) \Phi _{bl}(x)
\\
&+g_3\delta (x)e^{i\theta _3}u_{e1}+g_4\delta (x-d_b)e^{i\theta _4}u_{e2},
\\
\left. E\right. u_{e_1}=&\left( \omega_{e_1}-i\frac{\gamma _{e_1}}{2} \right) u_{e_1}+\Omega e^{i\alpha}u_{e_2}
\\
&+g_3e^{-i\theta _3}\delta (x)\left[ \Phi _{bR}(x)+\Phi _{bL}(x) \right] 
\\
&+g_2e^{-i\theta _2}\delta \left( x-d_a \right) \left[ \Phi _{aR}(x)+\Phi _{aL}(x) \right],
\\
\left. E \right. u_{e_2}=&\left( \omega_{e_2}-i\frac{\gamma _{e_2}}{2} \right) u_{e_1}+\Omega e^{-i\alpha}u_{e_1}
\\
&+g_1e^{-i\theta _1}\delta (x)\left[ \Phi _{aR}(x)+\Phi _{aL}(x) \right] 
\\
&+g_4e^{-i\theta _4}\delta \left( x-d_b \right) \left[ \Phi _{bR}(x)+\Phi _{bL}(x) \right]. 
\end{aligned}
\end{equation}
The wavefunctions $\Phi _{al,bl}(x)$ describe the various transport cases of propagating photons in the dual-rail \mbox{waveguide}. Without loss of generality, we assume that the photons are incident from the far left of the lower \mbox{waveguide}, the atoms are in the ground state $|g\rangle$, and the wave vectors of the propagating photons satisfy the linearized waveguide dispersion relation $E=w_0+k_{a,b}v_g$.

Substituting the wave function of the propagating photon in  Eqs.~(\ref{S14 wavefunction}) into Eqs.~(\ref{con:dynamics equation}), the scattering coefficient of the $\nabla$-type giant atom, as shown in Fig.~\ref{fig:scatterdetail}, can be solved as follows:
\begin{equation}
\label{con:S14 waveguide function group}
\begin{array}{l}
0=-iv_g\left( s_{1\rightarrow 1}-W_{a}^{L} \right) +g_1e^{i\theta _1}u_{e_2},
\\
0=-iv_g W_{a}^{L} e^{-i\phi _a}+g_2e^{i\theta _2}u_{e_1},
\\
0=-iv_g\left( s_{1\rightarrow 2}-W_{a}^{R} \right) e^{i\phi _a}+g_2e^{i\theta _2}u_{e_1},
\\
0=-iv_g( W_{a}^{R} -1)+g_1e^{i\theta _1}u_{e_2},
\\
0=-iv_g\left( s_{1\rightarrow 3}-W_{b}^{L} \right) +g_3e^{i\theta _3}u_{e_1},
\\
0=-iv_g W_{b}^{L} e^{-i\phi _b}+g_4e^{i\theta _4}u_{e_2},
\\
0=-iv_g W_{b}^{R} +g_4e^{i\theta _4}u_{e_1},
\\
0=-iv_g\left( s_{1\rightarrow 4}-W_{b}^{R} \right) e^{i\phi _b}+g_3e^{i\theta _3}u_{e_2}.
\end{array}
\end{equation}
We obtain
\begin{equation}
\label{AP_equation}
\begin{aligned}
0=&\left( \Delta+i\frac{\gamma _{e_1}}{2} \right) u_{e_1}-\Omega e^{i\alpha}u_{e_2}
\\
&-g_3e^{-i\theta _3}\left[ \frac{1}{2}W_{b}^{R}+\frac{1}{2} \left( W_{b}^{L} + s_{1\rightarrow3} \right) \right] 
\\
&-g_2e^{-i\theta _2} \left[ \frac{1}{2}e^{i\phi_a} \left( W_{a}^{R} + s_{1\rightarrow2} \right) + \frac{1}{2}e^{-i\phi_a}W_{a}^{L} \right],
\\
0=&\left( \Delta+i\frac{\gamma _{e_2}}{2} \right) u_{e_2}-\Omega e^{-i\alpha}u_{e_1}
\\
&-g_1e^{-i\theta _1}\left[ \frac{1}{2}(1+W_{a}^{R}) +\frac{1}{2}\left( W_{a}^{L} + s_{1\rightarrow1} \right) \right] 
\\
&-g_4e^{-i\theta _4} \left[ \frac{1}{2}e^{i \phi_b}(W_{b}^{R}+s_{1\rightarrow4})+\frac{1}{2}e^{-i\phi_b}W_{b}^{L} \right],
\end{aligned}
\end{equation}
where $\Delta=E-\omega_{e_{1,2}}=v_g k-\omega_{1,2}$ is the frequency detuning between the incident photon and the atomic level $|g\rangle \leftrightarrow |e_{1,2}\rangle $. Solving the above equation can calculate the photon scattering amplitudes $s_{1\rightarrow n}$ incident from port $1$. The scattering process of photons incident from ports 2, 3, and 4 can also be calculated based on the same principle. 


\section{Proposed experiment in superconducting circuits to realize a $\nabla$-type giant atom in a dual-rail waveguide}
\label{appendix:realization}

\begin{figure}[htp]
    \centering
    \includegraphics[width=0.9\linewidth]{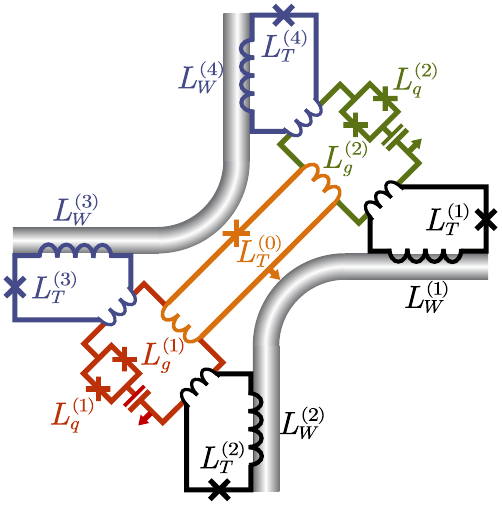}
    \caption{Implementation of the model using superconducting circuits.}
    \label{fig:exprealize}
\end{figure}

In our proposed setup, each coupling point between the giant atom and the waveguide is facilitated by a Josephson junction incorporated into a superconducting loop, as shown in Fig.~\ref{fig:exprealize}. The inductances $L_w^{(i)}$ and $L_q^{(i)}$ within the $i^{\text{th}}$ loop (with $i = 1, 2, 3, 4$) represent the shared branches connected to the waveguide and the giant atom, respectively. The gauge-invariant phase difference across the Josephson inductance in loop $i$ is denoted by $\phi_i$. The inductance $L_g^{(j)}$ corresponds to the interatomic loop of the $j^{\text{th}}$ qubit. The intermediate junction behaves as an effective lumped inductance $L_i$, expressed as
\begin{equation}
    L_i = \frac{L_{T}^{(i)}}{\cos \phi_i}, \quad L_{T}^{(i)} = \frac{\Phi_0}{2 \pi I_{c}^{(i)}},
\end{equation}
where $\Phi_0$ is the magnetic flux quantum, and $I_{c}^{(i)}$ denotes the critical current of the $i^{\text{th}}$ Josephson junction. When an external magnetic flux $\Phi_i^{\mathrm{ext}}$ is applied to loop $i$, the phase $\phi_i$ satisfies the equation~\cite{PhysRevA.92.012320}
\begin{equation}
    \phi_i + \beta_i \sin \phi_i = \frac{2 \pi}{\Phi_0} \Phi_i^{\mathrm{ext}}, \quad \beta_i = \frac{L_w^{(i)} + L_q^{(i)}}{L_{T}^{(i)}}.
\end{equation}
Using the $Y$-$\Delta$ transformation for the $i^{\text{th}}$ coupling loop, we derive the effective mutual inductance $M_i$ between the waveguide and the giant atom as~\cite{wulschner2016tunable}
\begin{equation}
    M_i = \frac{L_0^2}{2 L_0 + L_i} = \frac{L_0^2}{L_{T}^{(i)}} \frac{\cos \phi_i}{1 + \beta \cos \phi_i}.
\end{equation}

For simplicity, we assume identical inductances for all loops, setting $L_g^{(j)} = L_w^{(i)} = L_q^{(i)} = L_0$, with $L_0$ being much smaller than $L_{T}^{(i)}$ (i.e., $L_0 \ll L_{T}^{(i)}$). Under these assumptions, the mutual inductance simplifies to
\begin{equation}
    M_i = \frac{L_0^2}{L_{T}^{(i)}} \cos\left( \frac{2 \pi}{\Phi_0} \Phi_i^{\mathrm{ext}} \right).
\end{equation}
Similarly, the mutual inductance between the two giant atoms is given by
\begin{equation}
    M_g = \frac{L_0^2}{L_{T}^{(i)}} \cos\left( \frac{2 \pi}{\Phi_0} \Phi_g^{\mathrm{ext}} \right),
\end{equation}
where $\Phi_g^{\mathrm{ext}}$ is the external flux applied to the loop mediating the interaction between the two atoms.

The additional inductances contributed by the coupling loops to the two atoms are
\begin{align}
    L_a^{(1)} &= 2 L_0 + M_g + M_2 + M_3, \\
    L_a^{(2)} &= 2 L_0 + M_g + M_1 + M_4.
\end{align}
Including these additional inductances, the total inductance of the $j^{\text{th}}$ transmon qubit becomes~\cite{koch2007charge}
\begin{equation}
    L_Q^{(j)} = L_s^{(j)} + L_a^{(j)}, \quad L_s^{(j)} = \frac{\Phi_0^2}{E_J^{(j)}},
\end{equation}
where $L_s^{(j)}$ is the inductance of the SQUID loop in the transmon, and $E_J^{(j)}$ represents the Josephson energy of the $j^{\text{th}}$ qubit ($j = 1, 2$).

Considering that the transmon operates in a regime of weak Kerr nonlinearity, it can be approximated as a Duffing oscillator. The quantized Hamiltonian for the transmon is then
\begin{equation}
    H_q^{(j)} = \hbar \Omega_q^{(j)} b_j^\dagger b_j + \frac{E_C^{(j)}}{12} \left( b_j^\dagger + b_j \right)^4, \quad \Omega_q^{(j)} = \frac{1}{\sqrt{L_Q^{(j)} C^{(j)}}},
\end{equation}
where $C^{(j)}$ is the total capacitance of the $j^{\text{th}}$ transmon, $E_C^{(j)} = e^2/\left(2 C^{(j)}\right)$ is the charging energy, and $b_j$ ($b_j^\dagger$) are the annihilation (creation) operators for the transmon mode. Restricting the system to its two lowest energy levels, the Hamiltonian simplifies to
\begin{equation}
    H_q^{(j)} = \frac{1}{2} \hbar \omega_q^{(j)} \sigma_z^{(j)}, \quad \omega_q^{(j)} = \Omega_q^{(j)} - \frac{E_C^{(j)}}{\hbar}.
\end{equation}
The current operator for the transmon qubit is approximately~\cite{PhysRevA.92.012320}
\begin{equation}
    I_q^{(j)} = \sqrt{\frac{\hbar \omega_q^{(j)}}{2 L_Q^{(j)}}} \left( \sigma_-^{(j)} + \sigma_+^{(j)} \right),
\end{equation}
where $\sigma_-^{(j)}$ and $\sigma_+^{(j)}$ are the lowering and raising operators for the $j^{\text{th}}$ qubit.

The current operator for the waveguide at position $x$ is expressed as~\cite{wangTunableChiralBound2021}
\begin{equation}
    I_w(x) = \sum_{k} \sqrt{\frac{\hbar \omega(k)}{2 L_{\text{tot}}}} \left( a_k e^{i k x} - a_k^\dagger e^{-i k x} \right),
\end{equation}
where $a_k$ ($a_k^\dagger$) are the annihilation (creation) operators for the waveguide modes, and $L_{\text{tot}}$ is the total inductance of the waveguide.

The interaction Hamiltonian for the mutual inductance at the $i^{\text{th}}$ coupling point is then~\cite{wangChiralQuantumNetwork2022a}
\begin{equation}
\begin{split}
    H_{\mathrm{int}}^{(i)} &= M_i I_q^{(j)} I_w(x_i) \\
    &= \frac{L_0^2}{L_{T}^{(i)}} \cos\left( \frac{2 \pi}{\Phi_0} \Phi_i^{\mathrm{ext}} \right) 
    \sqrt{\frac{\hbar \omega_q^{(j)}}{2 L_Q^{(j)}}} \left( \sigma_-^{(j)} + \sigma_+^{(j)} \right) \\
    &\quad \times \sum_{k} \sqrt{\frac{\hbar \omega(k)}{2 L_{\text{tot}}}} 
    \left( a_k e^{i k x_i} - a_k^\dagger e^{-i k x_i} \right).
\end{split}
\end{equation}
Considering that only waveguide modes with frequencies close to $\omega_q^{(j)}$ interact significantly with the qubit, we can approximate the interaction Hamiltonian as~\cite{PhysRevA.99.063815}
\begin{equation}
\begin{split}
    H_{\mathrm{int}}^{(i)} &\approx \frac{L_0^2}{L_{T}^{(i)}} \cos\left( \frac{2 \pi}{\Phi_0} \Phi_i^{\mathrm{ext}} \right) 
    \sqrt{\frac{\hbar \omega_q^{(j)}}{2 L_Q^{(j)}}} \left( \sigma_-^{(j)} + \sigma_+^{(j)} \right) \\
    &\quad \times \sqrt{2 \pi} \sqrt{\frac{\hbar \omega_q^{(j)}}{2 L_{\text{tot}}}} \delta(x - x_i) 
    \left( a(x_i) + a^\dagger(x_i) \right),
\end{split}
\end{equation}
where $a(x)$ and $a^\dagger(x)$ are the field operators in real space, and $\delta(x - x_i)$ is the Dirac delta function, indicating that the interaction occurs at a specific point in space.

Similarly, the interaction Hamiltonian describing the coupling between the two qubits via their mutual inductance is
\begin{align}
    H_{\mathrm{int}}^{g} &= M_g I_q^{(1)} I_q^{(2)} \notag \\
    &= \frac{L_0^2}{L_{T}^{(i)}} 
    \cos\left( \frac{2 \pi}{\Phi_0} \Phi_g^{\mathrm{ext}} \right) 
    \sqrt{\frac{\hbar \omega_q^{(1)}}{2 L_Q^{(1)}}} \left( \sigma_-^{(1)} + \sigma_+^{(1)} \right) \notag \\
    &\quad \times \sqrt{\frac{\hbar \omega_q^{(2)}}{2 L_Q^{(2)}}} 
    \left( \sigma_-^{(2)} + \sigma_+^{(2)} \right).
\end{align}
This interaction facilitates direct coupling between the two qubits, enabling coherent energy exchange and entanglement.

By carefully designing the external flux biases $\Phi_i^{\mathrm{ext}}$ and $\Phi_g^{\mathrm{ext}}$, one can tune the mutual inductances $M_i$ and $M_g$, thereby controlling the strength of the interactions between the qubits and the waveguide, as well as between the qubits themselves. This tunability is crucial for implementing dynamic quantum control.


\section{Realization of Phase-Coupled Hamiltonian via Time-Dependent Coupling}
\label{appendix:phase_coupling}

In this section, we discuss how to achieve the desired phase-coupled Hamiltonian in our $\nabla$-type giant atom system by modulating the coupling coefficients in time. This approach allows us to introduce complex hopping terms analogous to those induced by an external gauge field, effectively implementing a synthetic magnetic flux in our circuit quantum electrodynamics (QED) setup.

\subsection{Time-Dependent Modulation of Coupling Strengths}

The mutual inductance between the giant atom and the waveguide at the $i^{\text{th}}$ coupling point is given by
\begin{equation}
    M_i = \frac{L_0^2}{L_{T}^{(i)}} \cos\left( \frac{2\pi}{\Phi_0} \Phi_i^{\mathrm{ext}} \right),
\end{equation}
where $L_0$ is the loop inductance, $L_T$ is the Josephson inductance, $\Phi_0$ is the magnetic flux quantum, and $\Phi_i^{\mathrm{ext}}$ is the external magnetic flux threading the $i^{\text{th}}$ coupling loop.

To introduce a controllable phase into the coupling, we modulate the external flux as a function of time~\cite{wangChiralQuantumNetwork2022a}
\begin{equation}
    \Phi_i^{\mathrm{ext}}(t) = \Phi_i^0 + \delta\Phi_i \cos(\Delta t + \theta_i),
\end{equation}
where $\Phi_i^0$ is the static flux bias, $\delta\Phi_i$ is the amplitude of the flux modulation, $\Delta$ is the modulation frequency, and $\theta_i$ is the initial phase offset.

Substituting $\Phi_i^{\mathrm{ext}}(t)$ into the expression for $M_i$, we obtain a time-dependent mutual inductance
\begin{equation}
    M_i(t) = \frac{L _0^2}{L_{T}^{(i)}} \cos\left( \frac{2\pi}{\Phi_0} \left[ \Phi_i^0 + \delta\Phi_i \cos(\Delta t + \theta_i) \right] \right).
\end{equation}

Assuming that $\delta\Phi_i$ is small compared to $\Phi_0$, we can expand the cosine function using the first-order Taylor expansion
\begin{equation}
    M_i(t) \approx M_i^0 - M_i^1 \cos(\Delta t + \theta_i),
\end{equation}
where
\begin{align}
    M_i^0 &= \frac{L_0^2}{L_{T}^{(i)}} \cos\left( \frac{2\pi}{\Phi_0} \Phi_i^0 \right), \\
    M_i^1 &= \frac{L_0^2}{L_{T}^{(i)}} \sin\left( \frac{2\pi}{\Phi_0} \Phi_i^0 \right) \left( \frac{2\pi}{\Phi_0} \delta\Phi_i \right).
\end{align}

\subsection{Abstract Model Correspondence}

Specifically, we consider the interaction between the giant atom and the waveguide at multiple coupling points with time-dependent mutual inductances $M_i(t)$. The Hamiltonian of our system can be expressed as
\begin{equation}
    H(t) = H_{\text{q}} + H_{\text{w}} + H_{\text{int}}(t),
\end{equation}
where $H_{\text{q}}$ is the Hamiltonian of the giant atom (qubit), $H_{\text{w}}$ is the Hamiltonian of the waveguide, and $H_{\text{int}}(t)$ represents the time-dependent interaction between the qubit and the waveguide.

A system of two qubits with time-dependent coupling can be utilized to engineer complex hopping terms~\cite{roushan2017chiral}. Analogously, in our circuit, the time-dependent mutual inductances $M_i(t)$ play the role of modulating the coupling between the qubit and the waveguide, allowing us to introduce controllable phases into the interaction terms.

\subsection{Derivation of Phase Coupling}

The interaction Hamiltonian at the $i^{\text{th}}$ coupling point is given by
\begin{equation}
    H_{\text{int}}^{(i)}(t) = M_i(t) I_{\text{q}}(t) I_{\text{w}}(x_i, t),
\end{equation}
where $I_{\text{q}}(t)$ is the current operator of the qubit, and $I_{\text{w}}(x_i, t)$ is the current operator of the waveguide at position $x_i$.

Expressing the qubit current operator in terms of the qubit lowering and raising operators,
\begin{equation}
    I_{\text{q}}(t) = I_{\text{q}} \left( \sigma_- e^{-i \omega_{\text{q}} t} + \sigma_+ e^{i \omega_{\text{q}} t} \right),
\end{equation}
and the waveguide current operator as~\cite{wangTunableChiralBound2021}
\begin{equation}
    I_{\text{w}}(x_i, t) = \int_{-\infty}^{+\infty} \frac{dk}{2\pi} I_k \left( a_k e^{i (k x_i - \omega_k t)} + a_k^\dagger e^{-i (k x_i - \omega_k t)} \right),
\end{equation}
we can write the interaction Hamiltonian as
\begin{align}
    H_{\text{int}}^{(i)}(t) &= \left[ M_i^0 - M_i^1 \cos(\Delta t + \theta_i) \right] I_{\text{q}} \left( \sigma_- e^{-i \omega_{\text{q}} t} + \sigma_+ e^{i \omega_{\text{q}} t} \right) \notag \\
    &\quad \times \int_{-\infty}^{+\infty} \frac{dk}{2\pi} I_k \left( a_k e^{i (k x_i - \omega_k t)} + a_k^\dagger e^{-i (k x_i - \omega_k t)} \right).
\end{align}

We now focus on the time-dependent part of the mutual inductance, as it enables us to introduce the desired phase coupling. Discarding the static term $M_i^0$ (which contributes to off-resonant interactions), we consider only the modulated part:
\begin{align}
    H_{\text{int}}^{(i)}(t) &= - M_i^1 \cos(\Delta t + \theta_i) I_{\text{q}} \left( \sigma_- e^{-i \omega_{\text{q}} t} + \sigma_+ e^{i \omega_{\text{q}} t} \right) \notag \\
    &\quad \times \int_{-\infty}^{+\infty} \frac{dk}{2\pi} I_k \left( a_k e^{i (k x_i - \omega_k t)} + a_k^\dagger e^{-i (k x_i - \omega_k t)} \right).
\end{align}

We rewrite the interaction Hamiltonian as
\begin{align}
    H_{\text{int}}^{(i)}(t) &= - \frac{M_i^1}{2} \left[ e^{i (\Delta t + \theta_i)} + e^{-i (\Delta t + \theta_i)} \right] I_{\text{q}} \\
    &\quad \times \left( \sigma_- e^{-i \omega_{\text{q}} t} + \sigma_+ e^{i \omega_{\text{q}} t} \right) \notag \\
    &\quad \times \int_{-\infty}^{+\infty} \frac{dk}{2\pi} I_k \left( a_k e^{i (k x_i - \omega_k t)} + a_k^\dagger e^{-i (k x_i - \omega_k t)} \right).
\end{align}

Combining exponential terms, the interaction Hamiltonian becomes
\begin{align}
    H_{\text{int}}^{(i)}(t) &= - \frac{M_i^1}{2} \Bigg\{ e^{i (\Delta t + \theta_i)} I_{\text{q}} \left( \sigma_- e^{-i \omega_{\text{q}} t} + \sigma_+ e^{i \omega_{\text{q}} t} \right) \notag \\
    &\quad \times  \int_{k} \frac{dk}{2 \pi} I_k \left( a_k e^{i (k x_i - \omega_k t)} + a_k^\dagger e^{-i (k x_i - \omega_k t)} \right) \notag \\
    &\quad + e^{-i (\Delta t + \theta_i)} I_{\text{q}} \left( \sigma_- e^{-i \omega_{\text{q}} t} + \sigma_+ e^{i \omega_{\text{q}} t} \right) \notag \\
    &\quad \times  \int_{k} \frac{dk}{2 \pi} I_k \left( a_k e^{i (k x_i - \omega_k t)} + a_k^\dagger e^{-i (k x_i - \omega_k t)} \right) \Bigg\}.
\end{align}

Next, we expand the products and collect terms oscillating at similar frequencies. The terms that contribute significantly are those that are near-resonant, satisfying energy conservation. We apply the rotating-wave approximation (RWA) and retain only the terms that vary slowly.

Assuming that the modulation frequency $\Delta$ satisfies
\begin{equation}
    \Delta = \omega_{\text{q}} - \omega_k,
\end{equation}
we can identify terms where the exponential time dependencies cancel out.

For example, consider the term
\begin{align}
    &\exp\{i (\Delta t + \theta_i)\} \sigma_- e^{-i \omega_{\text{q}} t} a_k \exp\{i (k x_i - \omega_k t)\} \notag \\
    &= \sigma_- a_k \exp\{i [(\Delta - \omega_{\text{q}} - \omega_k) t + k x_i + \theta_i]\}.
\end{align}

With $\Delta = \omega_{\text{q}} - \omega_k$, the time-dependent exponential becomes $e^{-i 2 \omega_{\text{q}} t}$, which oscillates rapidly and averages out over time. Therefore, we neglect such rapidly oscillating terms.

For example, a term that survives under the RWA is
\begin{align}
    &\exp\{-i (\Delta t + \theta_i)\} \sigma_+ e^{+i \omega_{\text{q}} t} a_k \exp\{i (k x_i - \omega_k t)\} \notag \\
    &= \sigma_+ a_k \exp\{-i [(\Delta - \omega_{\text{q}} + \omega_k) t - k x_i + \theta_i]\}.
\end{align}

With $\Delta = \omega_{\text{q}} - \omega_k$, the exponential becomes time-independent.

Collecting the relevant terms, the effective interaction Hamiltonian under the RWA becomes
\begin{equation}
    H_{\text{int}}^{(i)} = - \frac{M_i^1}{2} I_{\text{q}} I_k \left[ e^{-i \theta_i} \sigma_- a_k^\dagger e^{-i k x_i} + e^{i \theta_i} \sigma_+ a_k e^{i k x_i} \right].
\end{equation}

Transforming back to real space by integrating over $k$, we have
\begin{equation}
    H_{\text{int}}^{(i)} = g_i' \left[ e^{-i \theta_i} \sigma_- a^\dagger(x_i) + e^{i \theta_i} \sigma_+ a(x_i) \right],
\end{equation}
where
\begin{equation}
    g_i = - \frac{M_i^1}{2} I_{\text{q}} I_k.
\end{equation}

Thus, we obtain an effective interaction Hamiltonian with a controllable phase $\theta_i$ in the coupling terms. This phase arises directly from the initial phase of the flux modulation $\theta_i$, allowing us to engineer complex hopping terms in our system.

A similar analysis can be applied to the interatomic interaction, applying a time-modulated external flux $\Phi_g^{\mathrm{ext}}(t) = \Phi_g^0 + \delta\Phi_g \cos(\Delta t + \alpha)$, the interatomic strength is
\begin{equation}
    \Omega = -\frac{M_{g}^{1}}{2} I_{q}^{(1)} I_{q}^{(2)}. 
\end{equation}




\bibliography{ref.bib} 

\end{document}